\documentclass[preprint,pra,showpacs,showkeys,floatfix]{revtex4}

\baselineskip 
\usepackage{amsmath}
\usepackage{amsfonts}            
\usepackage{amssymb}    

\usepackage{graphicx}

\begin{document}

\title{Topology of surfaces for molecular Stark energy, alignment and orientation
generated by combined permanent and induced electric dipole interactions
}

\author{Burkhard Schmidt}

\email{burkhard.schmidt@fu-berlin.de}

\affiliation{%
Institute for Mathematics, Freie Universit\"{a}t Berlin \\ Arnimallee 6, D-14195 Berlin, Germany}%

\author{Bretislav Friedrich}

\email{brich@fhi-berlin.mpg.de}

\affiliation{%
Fritz-Haber-Institut der Max-Planck-Gesellschaft \\ Faradayweg 4-6, D-14195 Berlin, Germany~}%

\date{\today}

\begin{abstract}

We show that combined permanent and induced electric dipole interactions of polar and polarizable molecules with collinear electric fields lead to a {\it sui generis} topology of the corresponding Stark energy surfaces and of other observables -- such as alignment and orientation cosines -- in the plane spanned by the permanent and induced dipole interaction parameters. We find that the loci of the intersections of the surfaces can be traced analytically and that the eigenstates as well as the number of their intersections can be characterized by a single integer index.  The value of the index,  distinctive for a particular ratio of the interaction parameters, brings out a close kinship with the eigenproperties obtained previously for a class of Stark states via the apparatus of supersymmetric quantum mechanics.

\end{abstract}

\maketitle

\section{Introduction}

The pursuit of means to manipulate molecular rotation and translation is a leading frontier of chemical/molecular physics. 
Among recent developments are new methods to control the orientation and/or alignment of molecules \cite{SlenFriHerPRL1994,JCP1999FriHer, FriHerJPCA99, Ortigoso1999,Seideman:99a,Seideman1999,Larsen:00a,CaiFri2001,Averbukh:01a, Leibscher:03a,PRL2003Sakai, JMO2003-Fri,Leibscher:04a,Buck-Farnik,HaerteltFriedrichJCP08,PRL2009Stap, Owschimikow2009,Ohshima2010,Averbukh2010,Owschimikow2010,Owschimikow2011} as well as methods to deflect and focus their translational motion \cite{JCP2009Kuepp} and to achieve molecular trapping \cite{ChemRev2012-Meijer}. 
The importance of orientation comes also to light in novel applications such as attaining time-resolved photoelectron angular distributions \cite{Holmegaard2010,Bisgaard2009,Hansen2011}, diffraction-from-within \cite{Landers2001}, separation of photodissociation products \cite {IsraelJ2003-Manz,JCP2004-Manz, Lorenz:11a}, deracemization \cite {JCP2009Stapelfeldt}, high-order harmonic generation and orbital imaging \cite{Itatani:04a,CorkumHHG, BandraukHHG, IvanovHHG, Smirnova2009,Woerner,Mohn2012}, quantum simulation \cite{Bar:12,Man:13} or quantum computing \cite{DeMille2002,de6,de7,de8,QiKaisFrieHer2011a,QiKaisFrieHer2011b}.
All methods to manipulate molecular trajectories rely on the ability to create directional states of molecules. 
This is because only in directional states are the molecular body-fixed multipole moments ``available'' in the laboratory frame where they can be acted upon by space-fixed fields. 
In the case of polar molecules, the body-fixed permanent dipole moment is put to such a full use in the laboratory by creating oriented states characterized by as complete a projection of the body-fixed dipole moment on the space-fixed axis as the uncertainty principle allows. 
Such a high degree of orientation can now be achieved by a versatile technique  \cite{JCP1999FriHer,JCP2003-Buck,JMO2003-Fri,HaerteltFriedrichJCP08,Poterya:08} that combines a static electric field with a nonresonant optical field. 
The combined fields give rise to an amplification effect which occurs for any polar molecule, as only an anisotropic polarizability, along with a permanent dipole moment, is required. 
This is always available in polar molecules. 
Thus, for a variety of molecules in their rotational ground state, a very weak static electric field can convert second-order alignment by a laser into a strong first-order orientation that projects up to 90\% of the body-fixed dipole moment on the static field direction.  
The ``combined fields'' technique has found applications ranging from molecular imaging to surface science~\cite{StapelfeldtSeidemanRMP03, KremsPCCP08, StapPCCP2011,StapPRL2012,VattuoneProgSurfSci10} and has  been extended  to the case of molecules trapped in octahedral crystal fields \cite{Kiljunen:05a,Kiljunen:05b,Kiljunen2006}. 

\begin{figure}[htbp]
\centering
\includegraphics[width=5cm]{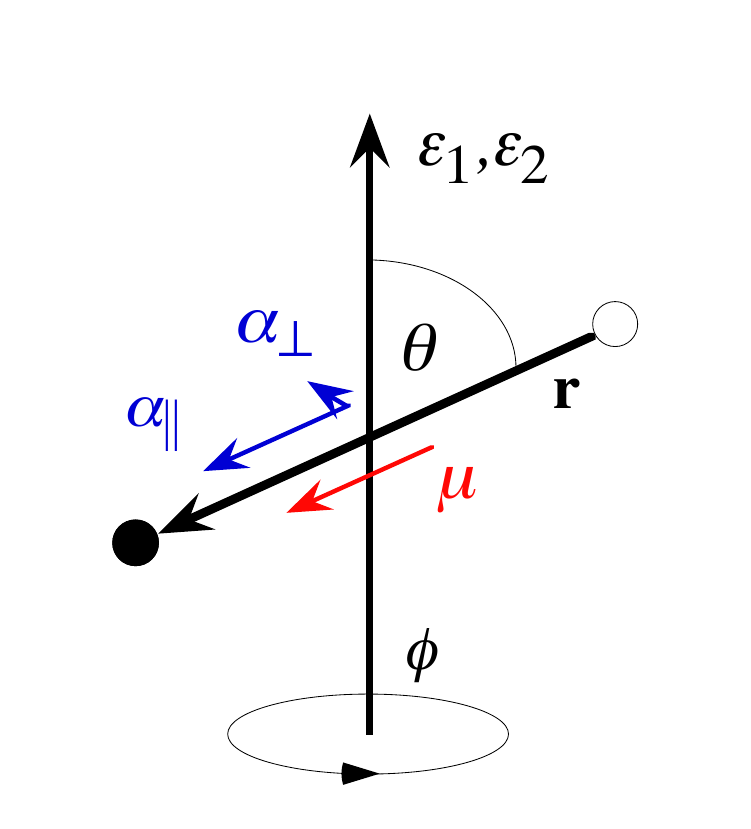}
\caption{\label{fig:molecule-field} {\small 
Collinear electric fields, $\boldsymbol{\varepsilon}_1$ and $\boldsymbol{\varepsilon}_2$, acting on the molecular dipole moment, $\mu$, and the parallel, $\alpha _{|| }$, and perpendicular, $\alpha _{\bot }$, components of the molecular polarizability. Also shown are the polar angle $\theta$ between the common direction of the field vectors and the direction of the molecular axis, $r$, as well as the uniformly distributed azimuthal angle, $\phi$ about the field vector.}}
\end{figure}

In our previous work, the permanent and induced dipole interactions  were assumed to arise, respectively, from electrostatic and nonresonant optical fields whose strength could be varied independently of each other, with the induced dipole interaction dominating over the permanent dipole interaction.
Herein we investigate aspects of the combined interactions of a polar and polarizable molecule with either different collinear fields or the same field that span the whole range of interaction strengths for both interactions.

The combined permanent and induced dipole interactions lead to a {\it sui generis} topology of the corresponding Stark energy surfaces and other observables spanned by the permanent and induced dipole interaction parameters, with intersections whose loci can be traced analytically. The eigenstates as well as the number of their intersections can be characterized by a single index whose value, distinctive for a particular ratio of the interaction parameters, brings out a close kinship with the eigenproperties obtained previously for a class of Stark states via the apparatus of supersymmetric quantum mechanics \cite{LemMusKaisFriPRA2011,LemMusKaisFriNJP2011}. Although the present work deals with eigenproperties, it prepares the soil for our forthcoming work on the dynamics of directional states of polar and polarizable molecules created by the inherently non-adiabatic interaction \cite {Ortigoso1999,CCCC2001} with a half-cycle pulse of a nonresonant optical field \cite{Bucksbaum1993, Jauslin2005,KrauszIvanov2009}. Such a pulse gives rise to both the permanent and induced dipole interactions at the same time.

This paper is organized as follows. In Sec. II, we briefly describe the Hamiltonian of a polar and polarizable molecule subject to collinear fields as a function of reduced dimensionless parameters that characterize the strengths of the permanent and induced dipole interactions. In Sec. III we present the eigenproperties of the above Hamiltonian such as eigenenergies, energy gaps between adjacent levels and directional properties (orientation and alignment cosines), and characterize the topology of the dependence of these eigenproperties on the reduced dimensionless parameters as a function of their ratio. In Sec. IV, we apply our results to the case where both the permanent and induced dipole interactions arise from the same field, in which case the ratio of the permanent to induced-dipole interaction is fixed for a given molecule. Finally, we discuss the ramifications of our findings for the dynamics of states created by time-dependent fields and point out a connection of the topology of the Stark energy surfaces to supersymmetry. 
 
\section{Theory}

The Hamiltonian of a polar $^{1}\Sigma $ rigid-rotor molecule with a
body-fixed dipole moment $\mu $, body-fixed static-polarizability components 
$\alpha _{||}$ and $\alpha _{\bot }$, and a rotational constant $B$ subject
to collinear electric fields $\boldsymbol{\varepsilon }_1$ and $\boldsymbol{\varepsilon }_2$ is given by 
\begin{equation}
H=B\mathbf{J}^{2}+V_{\mu }+V_{\alpha }  \label{hamiltonian}
\end{equation}
where $\mathbf{J}^{2}$ is the operator of square angular momentum, 
\begin{equation}
V_{\mu }=-\mu \varepsilon_1 \cos \theta
\end{equation}
\begin{equation}
V_{\alpha }=-\frac{1}{2}(\alpha _{||}-\alpha _{\bot })\varepsilon_2
^{2}\cos ^{2}\theta -\frac{1}{2}\alpha _{\bot }\varepsilon_2 ^{2}
\end{equation}
are, respectively, the permanent- and induced-dipole moment
potentials, $\varepsilon_{1,2} \equiv |\boldsymbol{\varepsilon}_{1,2}|$ are the
electric field strengths acting on the permanent and induced dipole moments, respectively, and $\theta $ is the polar angle between the common direction of $\boldsymbol{\varepsilon}_{1}$ and $\boldsymbol{\varepsilon}_{2}$ and the direction of the
molecular axis, ${\bf r}$, see Figure \ref{fig:molecule-field}. We note that $\boldsymbol{\varepsilon}_{1}$ can be due to an electrostatic field and $\boldsymbol{\varepsilon}_{2}$ to a non-resonant optical field of intensity $I$ such that 
\begin{equation}
\varepsilon_2 =\left( \frac{2 I}{c\epsilon_0}\right) ^{1/2}
\end{equation}
with $c$ the speed of light in vacuum and $\epsilon_0$ the vacuum permittivity. In this case, the fields $\boldsymbol{\varepsilon}_{1}$ and $\boldsymbol{\varepsilon}_{2}$ would indeed act on the permanent and induced dipoles separately, without adding up to a single effective field. However, the induced and permanent dipole interactions can also arise due to the same field $\boldsymbol{\varepsilon}_{1}=\boldsymbol{\varepsilon}_{2}=\boldsymbol{\varepsilon}$, in which case the two interactions maintain a fixed ratio for a given molecule, as will be discussed in Sec. V.

The Hamiltonian, eq. (\ref{hamiltonian}), can be recast in dimensionless
form by dividing through the rotational constant $B$; as a result 
\begin{equation}
\frac{H}{B}=\mathbf{J}^{2}-\eta \cos \theta -\Delta \eta
\cos ^{2}\theta -\eta _{\bot }  \label{redham}
\end{equation}
where 
\begin{equation}
\eta \equiv \frac{\mu \varepsilon_1}{B}\text{\hspace{1cm}}\Delta \eta \equiv \eta
_{||}-\eta _{\bot }\text{\hspace{1cm}}\eta _{||,\bot }\equiv \frac{%
\alpha _{||,\bot }\varepsilon_2^{2}}{2B}.
\label{defparam}
\end{equation}
We note that  the interaction strength is characterized by
the parameters $\eta $ and $\Delta \eta $ for any $^1\Sigma$ molecule. The eigenproperties obtained from
the reduced  Schr\"{o}dinger equation 
\begin{equation}
\frac{H}{B}\Psi=\frac{E}{B}\Psi
\label{redSE}
\end{equation}
are thus arbitrarily
``transferrable'' from one molecular species to another. Table \ref{table:parameters} lists the molecular parameters for a set of representative $^1\Sigma$ molecules (along with a couple of other symmetry species). Conversion factors needed to obtain the dimensionless reduced parameters from the molecular parameters expressed in terms of customary units are given in Table \ref{table:conversion}.

\begin{table}[h]
\centering
\caption{\small Parameters for representative linear molecules, see text. Compilation based on Refs. \cite{DulieuJCP2005,DulieuJCP2008} for the alkali dimers, Ref. \cite{Deigl2013Priv} for HD, and on Ref. \cite {FriHerJPCA99} for the rest. }
\vspace{.3cm}
\label{table:parameters}
\begin{tabular}{| l | c | c | c | c | c | c | c | c |}
\hline 
\hline
Molecule & $B$ [cm$^{-1}$] & $\mu$ [D] & $\Delta \alpha $ [\AA$^3$] & $\Delta \alpha$ [\AA$^3$]$B$ [cm$^{-1}$]/$\mu^2$ [D]  & $\Delta \eta/\eta^2$ & $k$  \\[5pt]
\hline

CsF(X$^1\Sigma$) & 0.1843  &  7.87 & (3.0) & 8.93$\times 10^{-3}$ & 8.83$\times 10^{-7}$ & 532.0   \\[5pt]

ICN(X$^1\Sigma$) &  0.1075  &  3.72 & (7.0) & 5.44$\times 10^{-2}$ & 5.38$\times 10^{-6}$ & 215.6   \\[5pt]

LiCs(X$^1\Sigma$) &  0.188 &  5.52 & 49.5 & 3.07$\times 10^{-1}$ & 3.04$\times 10^{-5}$ & 90.6   \\[5pt]

NaK(X$^1\Sigma$) &  0.091 &  2.76 & 39.5 & 4.72$\times 10^{-1}$ & 4.67$\times 10^{-5}$ & 73.1   \\[5pt]

KCs(X$^1\Sigma$) &  0.033 &  1.92 & 64.6 & 5.8$\times 10^{-1}$ & 5.72$\times 10^{-5}$ & 66.1   \\[5pt]

RbCs(X$^1\Sigma$) &  0.016 &  1.27 & 72.8 & 7.22$\times 10^{-1}$ & 7.14$\times 10^{-5}$ & 59.1   \\[5pt]

ICl(X$^1\Sigma$) &  0.1142  &  1.24  & (9.0) & 6.68$\times 10^{-1}$ & 6.61$\times 10^{-5}$ & 61.5   \\[5pt]

CO(A$^3\Sigma$) &  1.681  &  1.37    & (1.5) & 1.34 & 1.33$\times 10^{-4}$ & 43.3  \\[5pt]

OCS(X$^1\Sigma$) &  0.2039  &  0.709 & 4.1 & 1.66 & 1.64$\times 10^{-4}$ & 39.0   \\[5pt]

KRb(X$^1\Sigma$) &  0.032 &  0.76 & 54.1 & 2.99 & 2.96$\times 10^{-4}$ & 29   \\[5pt]

LiNa(X$^1\Sigma$) &  0.38 &  0.566 & 24.7 & 29.29$\times 10^{1}$ & 2.89$\times 10^{-3}$ & 9.3   \\[5pt]

NO(X$^2\Pi$) &  1.703  &  0.16    & 2.8 & 1.86$\times 10^{2}$ & 1.82$\times 10^{-2}$ & 3.7   \\[5pt]

CO(X$^1\Sigma$) &  1.931  &  0.10 & 1.0 & 1.93$\times 10^{2}$ & 1.91$\times 10^{-2}$ & 3.6  \\[5pt]

HD(X$^1\Sigma$) &  45.644  &  5$\times 10^{-4}$ & 0.305 & 5.56$\times 10^{7}$ & 5.508$\times 10^{3}$ & 6.7$\times 10^{-3}$  \\[5pt]

 \hline
 \hline
  
\end{tabular}
\end{table}

The eigenproperties of Hamiltonian (\ref{redham}) were obtained by expanding its eigenfunctions $\Psi$ in the free-rotor basis set, $|J,M\rangle$, 
\begin{equation}
	\label{PendularState}
	\Psi  = \sum_{J} c_{J M}^{\tilde{J}, M} (\eta, \Delta \eta) |J,M\rangle \equiv  |\tilde{J}, M;\eta, \Delta \eta\rangle
\end{equation}
and diagonalizing the corresponding Hamiltonian matrix truncated at $J_{\text {max}}=100$, which sufficed to achieve convergence for the range of the field strengths considered.  
The hybridization coefficients $c_{J M}^{\tilde{J}, M} (\eta, \Delta \eta)$ depend, for a given state $|\tilde{J}, M;\eta, \Delta \eta\rangle$, solely on the reduced interaction parameters, as indicated. We note that the projection, $M$, of the angular momentum $\mathbf{J}$ on $\boldsymbol{\varepsilon}_{1,2}$ is a good quantum number while $J$ is not. 
However, the value of $J$ of the field-free rotational state $|J,M\rangle$ that adiabatically correlates with the hybrid state can be used as a label, which we designate by $\tilde{J}$, so that $|\tilde{J}, M;\eta, \Delta \eta \rangle \to |J,M\rangle$ for $\eta, \Delta \eta \to 0$. Below we present results for states with $M=0$, which render a playing field large-enough to capture the salient features of the combined-fields problem's topology. 
In what follows, we'll simplify our notation and label the $|\tilde{J}, M=0;\eta, \Delta \eta\rangle$ states by $|\tilde{J}\rangle$. 
For $\eta>0$, the states $|\tilde{J}\rangle$ have an indefinite parity. We note that for $\eta=0$, the states have a definite parity, given by $(-1)^{\tilde{J}}$, independent of the value of $\Delta \eta$.

\begin{table}[h]
\centering
\caption{\small Conversion factors needed to obtain the dimensionless reduced parameters from the molecular parameters expressed in terms of customary units.}
\vspace{.3cm}
\label{table:conversion}
\begin{tabular}{| c | c | }
\hline 
\hline
Parameter & Expression  \\[5pt]
\hline
$\eta$ & $1.68\times10^{-2}\varepsilon$[kV/cm] $\mu$[Debye]/$B$[cm$^{-1}$] \\[5pt]
$\Delta\eta$ &  $2.79\times10^{-8}\varepsilon^2$[kV/cm] $\Delta\alpha$[\r{A}$^3$]/$B$[cm$^{-1}$]\\[5pt]
$\frac{\Delta \eta}{\eta^2}$ &  $9.892\times10^{-5} \eta^2 \Delta \alpha$[\r{A}$^3$] $B$[cm$^{-1}$]/$\mu^2$[Debye] \\[5pt]

\hline
\hline
  
\end{tabular}
\end{table}

\section{Results and Discussion}

\begin{figure}[htbp]
\includegraphics[width=07.5cm]{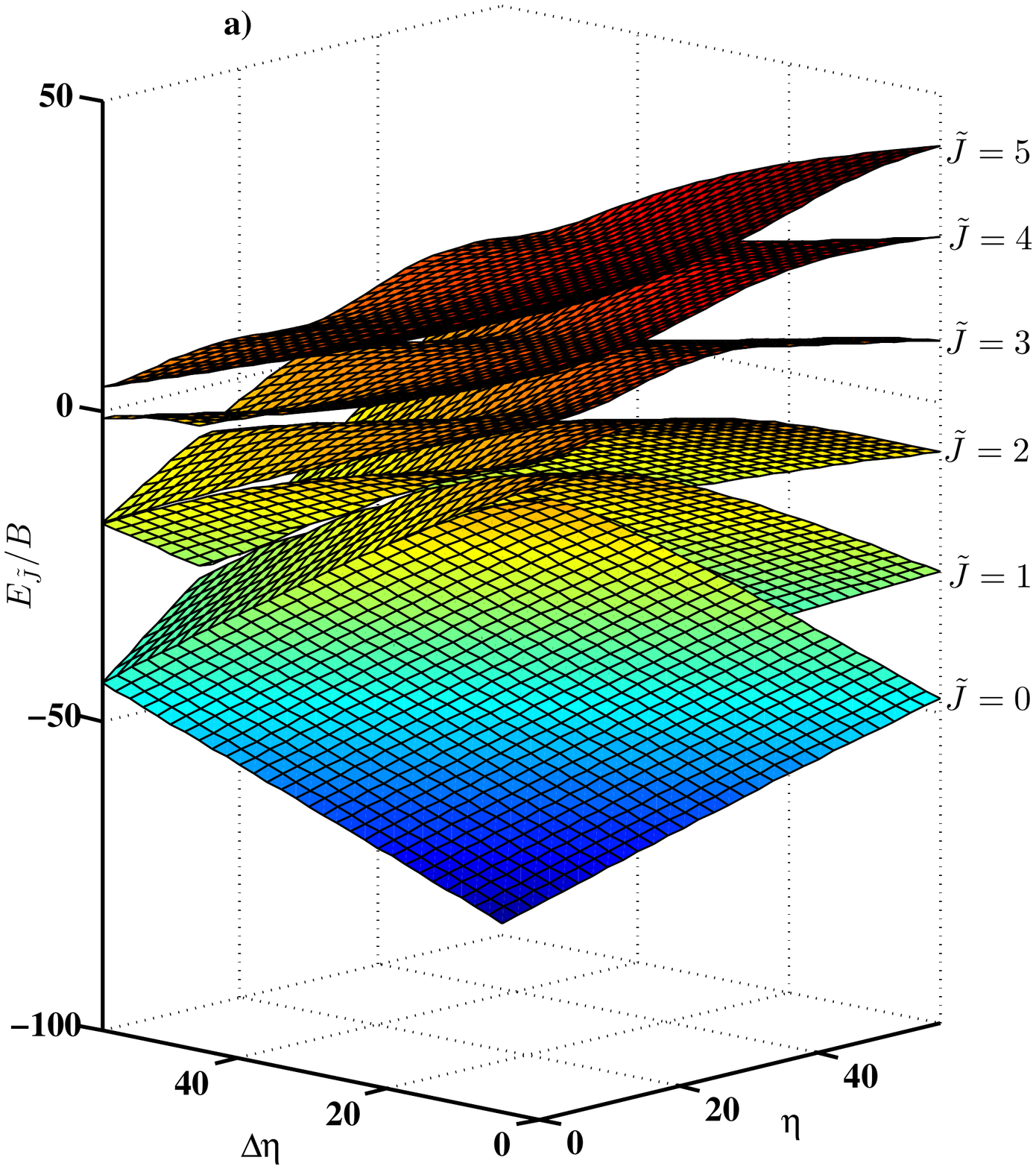}
\includegraphics[width=07.5cm]{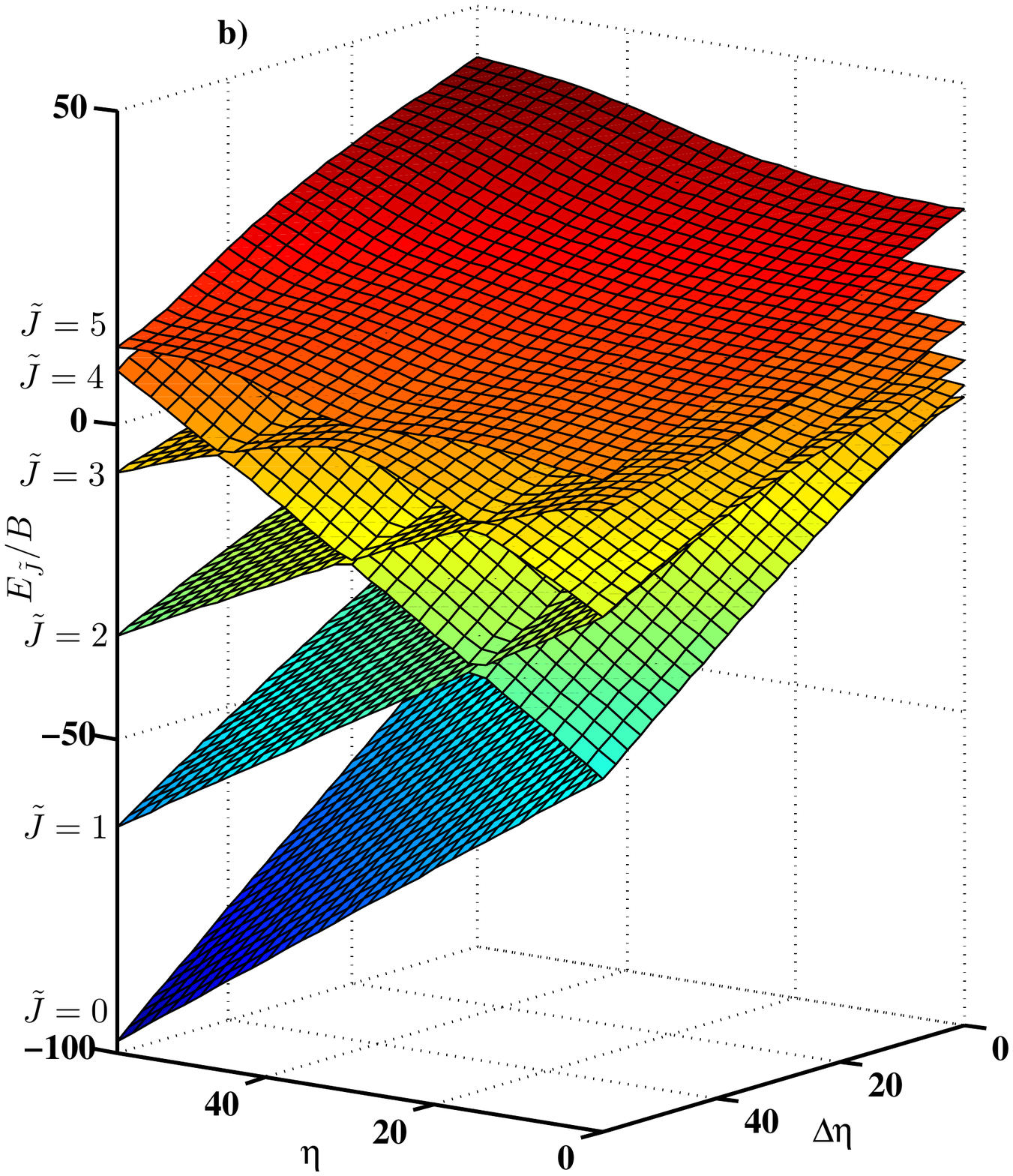}
\caption{Views of the lowest six reduced energy surfaces, $E_{\tilde{J}}(\eta,\Delta\eta)/B$, of  Hamiltonian (\ref{redham}) for a linear molecule subject to an electric field. The Stark energy surfaces are shown as functions of the parameters $\eta$ and $\Delta\eta$ that characterize the strengths of, respectively, the permanent and induced dipole interactions.}
\label{fig:pes}
\end{figure}

Fig. \ref{fig:pes} shows the resulting Stark energy surfaces pertaining to the lowest six eigenstates as functions of the parameters $\eta$ and $\Delta\eta$ that characterize the strengths of the permanent and induced dipole interactions.
In order to rationalize the observed features of the displayed energy surfaces, we first consider the case when the molecule interacts solely via the permanent dipole interaction, i.e., $\Delta \eta =0$, which roughly corresponds to the case of a polar molecule subject to a weak electrostatic field. 
As can be seen in Fig. \ref{fig:pes}a, the energy of the ground state, $\tilde{J}=0$, monotonously decreases with increasing $\eta$ (i.e., the state is  high-field seeking). In marked contrast, the eigenenergies of all the other states first increase with increasing $\eta$, run through an inflection point at $E/B=\eta$ (where the given state just becomes bound), and reach a maximum at $\eta\approx 2.15 \tilde{J}(\tilde{J}+1)+1.2$, beyond which the eigenenergies decrease again, without undergoing any curve crossings or exhibiting degeneracies. 

When a  molecule interacts solely via its induced dipole moment, i.e., when $\eta=0$ as would be the case for a non-polar molecule in an electrostatic or many-cycle non-resonant optical field, the eigenenergies monotonously decrease with the increasing interaction parameter $\Delta\eta$ (the states are all high-field seeking), see Fig.  \ref{fig:pes}b. Adjacent states, $|\tilde{J}\rangle$ and $|\tilde{J}+1\rangle$ with $\tilde{J}$ even, have opposite parity and form tunneling doublets. 
The interaction strength at which the doublet splitting drops below $B$ is $\Delta\eta\approx-2.6(\tilde{J}+1)^2-9.1(\tilde{J}+1)+14$.
The splitting of the tunneling doublets decreases as \cite{JCP1999FriHer, FriHerJPCA99}
\begin{equation}
\Delta E/B\equiv (E_{{\tilde{J}}+1}-E_{{\tilde{J}}})/B\propto \exp[-\Delta \eta^{\frac{1}{2}}] 
\label{eqn:split}
\end{equation} 
rendering the members of a given tunneling doublet quasi-degenerate and drops to zero altogether 
in the high field regime, where the interaction approaches the harmonic librator limit. There, the eigenenergies of the tunneling doublets are given by~\cite{FriHerZPhys96, FriHerJPC95}
\begin{equation}
\begin{split}
	      E_{{\tilde{J}}}/B=  - \Delta \eta  + 2\tilde{J} \Delta \eta^{\frac{1}{2}} +2 \Delta \eta^{\frac{1}{2}} - \frac{{\tilde{J}}^2}{2} - {\tilde{J}} - 1 \\
	       = - \Delta \eta + 2({\tilde{J}}+1)  \Delta \eta^{\frac{1}{2}} - \frac{({\tilde{J}}+1)^2}{2} - \frac{1}{2}=E_{{\tilde{J}}+1}/B \\
	       \hspace{0.3cm} \text{with ${\tilde{J}}=2n$ and $n=0,1,2, ...$}
	       \label{eqn:StrongFieldE}
\end{split}
\end{equation}
from which it follows that the reduced energy difference between the $(\tilde{J}/2)$-th doublet and the $(k+\tilde{J}/2)$-th doublet 
\begin{equation}
E_{{\tilde{J}}+2k}/B-E_{{\tilde{J}}}/B=4k\Delta\eta^{\frac{1}{2}}-2(\tilde{J}+1)k -2k^2 
\label{eqn:doublets}
\end{equation} 
We note that the gap between adjacent tunneling doublets (such as $|0\rangle$,$|1\rangle$ and $|2\rangle$,$|3\rangle$) becomes
\begin{equation}
E_{{\tilde{J}}+2}/B-E_{{\tilde{J}}}/B=4\Delta\eta^{1/2}-2({\tilde{J}}+2) 
\label{eqn:adjacentdoublets}
\end{equation} 
This energy separation between adjacent quasi-degenerate tunneling doublets as well as the tunneling splitting of Eq. (\ref{eqn:split}) become accurate for, e.g., the two lowest doublets at  $\Delta \eta > 50$ \cite{FriHerZPhys96}. 

\begin{figure}[htbp]
\includegraphics[width=8cm]{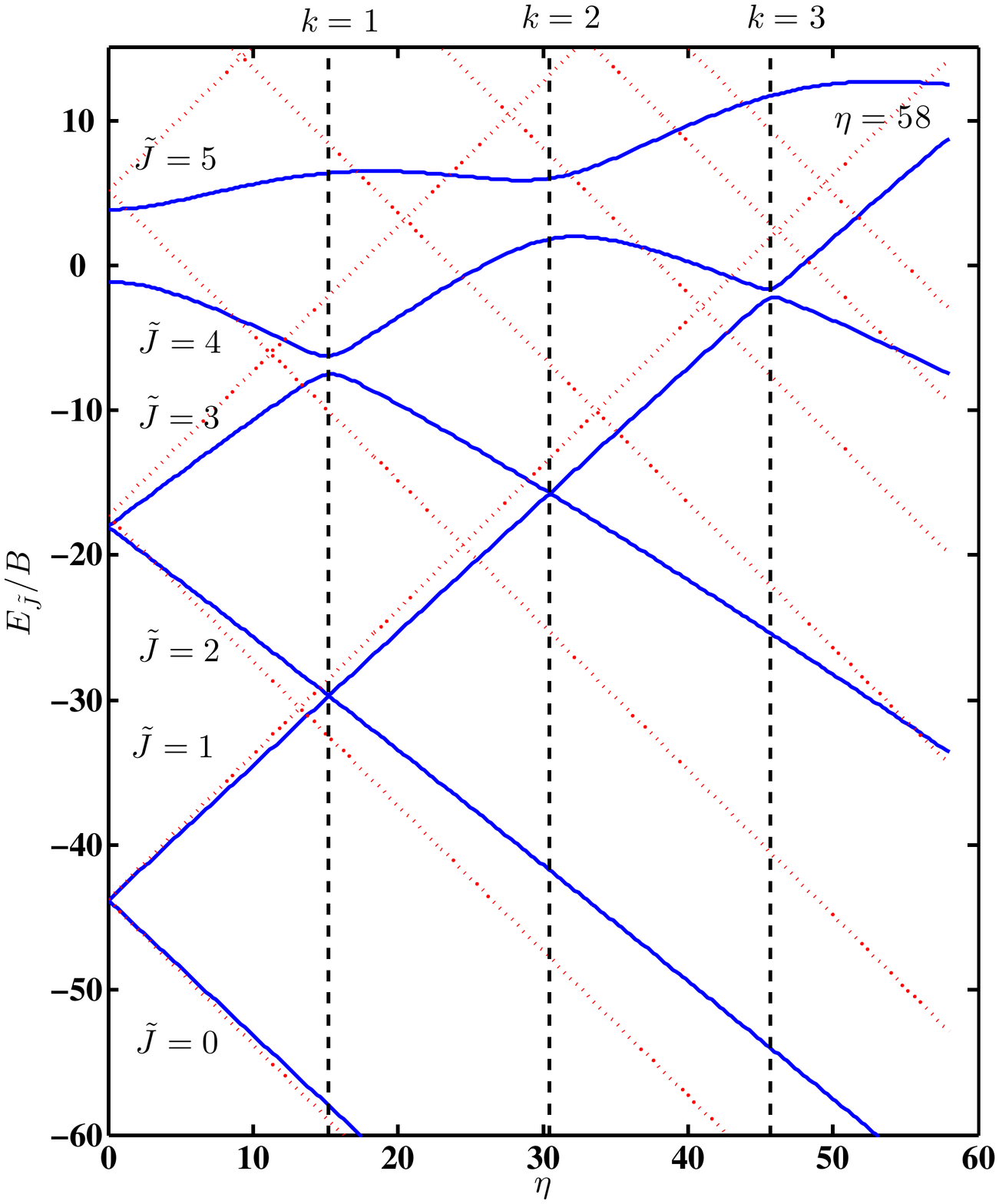}
\caption{Reduced eigenenergies, $E_{\tilde{J}}/B$, of a linear molecule subject to an electric field for $\Delta \eta=58$ as a function of $\eta$. Note that this value of $\Delta \eta$ connects to Fig. \ref{fig:pes} where  $\Delta \eta$ (and $\eta$) range up to the value of 58. Blue curves: Numerically obtained energies. Red lines: Energies in the harmonic librator limit assuming linear $\eta$--dependence of eq. (\ref{eqn:eequal}), with the loci of their intersections given by eq. (\ref{eqn:eta-Deta}). Black vertical lines: Intersection loci due to eq. (\ref{eqn:eta-Deta-simple}). }
\label{fig:intersections}
\end{figure}

As shown in our previous work on polar and polarizable molecules subject to combined static and optical fields \cite{JCP1999FriHer, FriHerJPCA99}, for a large-enough induced dipole interaction that renders the members of a given tunneling doublet quasi-degenerate, a very weak permanent dipole interaction, $\eta \ll \Delta\eta$, is sufficient to couple the opposite-parity members of the tunneling doublets and thus create highly oriented states (of indefinite parity). 
By making use of a two-state model \cite{FriHerJPCA99}, we were able to show  that
the energy levels  of the members of the tunneling doublets repel each other approximately proportionately to the strength $\eta$ of the permanent dipole interaction, see also the energy surfaces in Fig. \ref{fig:pes}b and the red lines in Fig.~\ref{fig:intersections} which show schematically the energies as a function of $\eta$ for fixed $\Delta \eta=200$ in the harmonic librator limit. 
For a large-enough permanent dipole interaction, this leads to a hierarchy of intersections between the $(\tilde{J}+1)$-th state and the $(\tilde{J}+2k)$-th state, i.e., between the upper member of the $(\tilde{J}/2)$-th doublet and the lower member of the $(\tilde{J}/2+k)$-th doublet (with $\tilde J$ even, cf. eq. (\ref{eqn:StrongFieldE})). 
Within the linear approximation of Ref.~\cite{FriHerJPCA99} for energy splittings of the tunneling doublets with $\eta$ at a given $\Delta \eta$, these intersections occur at energies
\begin{equation}
E_{\tilde{J}+1}/B+\eta=E_{\tilde{J}+2k}/B-\eta
\label{eqn:eequal}
\end{equation}
which, upon substitution from eq. (\ref{eqn:adjacentdoublets}), yields 
\begin{equation}
\Delta \eta = \frac{1}{4k^2} (\eta+(\tilde{J}+1)k+k^2)^2
\label{eqn:Deta-eta}
\end{equation}
or, equivalently,
\begin{equation}
\eta = 2k\Delta\eta^{1/2}-(\tilde{J}+1)k-k^2   \hspace{0.3cm} \text{with ${\tilde{J}}=2n$, $n=0,1,2, ...,$ and $k=1,2,3, ...$}
\label{eqn:eta-Deta}
\end{equation}
These intersection points are visible as the crossings of the red lines in Fig. \ref{fig:intersections} which correspond to the energies in the harmonic librator limit as a function of $\eta$ for fixed $\Delta \eta=58$, assuming a linear $\eta$--dependence employed in eq.~(\ref{eqn:eequal}).
Fig. \ref{fig:intersections}  also shows, by blue curves, the numerically obtained Stark energy surfaces for the combined-fields system.
Remarkably, the loci of their intersections are found at values
\begin{equation}
\eta = 2k\Delta\eta^{1/2}
\label{eqn:eta-Deta-simple}
\end{equation}
which are indicated by vertical lines in Fig. \ref{fig:intersections}.
Note that at values of ($\eta, \Delta \eta$) well below the harmonic librator limit, these values are not too far from the loci obtained in the harmonic librator limit, eq.~(\ref{eqn:eta-Deta}).  

As eq.~(\ref{eqn:eta-Deta-simple}) is independent of $\tilde{J}$, the number of intersections an energy surface partakes in is equal to the adiabatic label $\tilde{J}$ of the corresponding eigenstate: 
the lowest energy surface, with $\tilde{J}=0$, is thus not involved in any intersection; 
the first excited state surface, with $\tilde{J}=1$, is involved in a first-order ($k=1$) intersection (between nearest doublets); 
the second excited state surface, with $\tilde{J}=2$, is involved both in a first-order ($k=1$) intersection (between nearest doublets) and in a second-order ($k=2$) intersection (between second nearest doublets), etc. Consequently, at the loci of the $k$-th order intersections given by eq.~(\ref{eqn:eta-Deta-simple}), we find an energy level pattern with $k$ single states at the bottom, followed by all other states which are doubly degenerate. In contrast, there are no degeneracies arising anywhere between the intersection loci, as can be seen in both Fig.~\ref{fig:intersections} and Fig.~\ref{fig:pes}a. 

\begin{figure}[htbp]
\includegraphics[width=12cm]{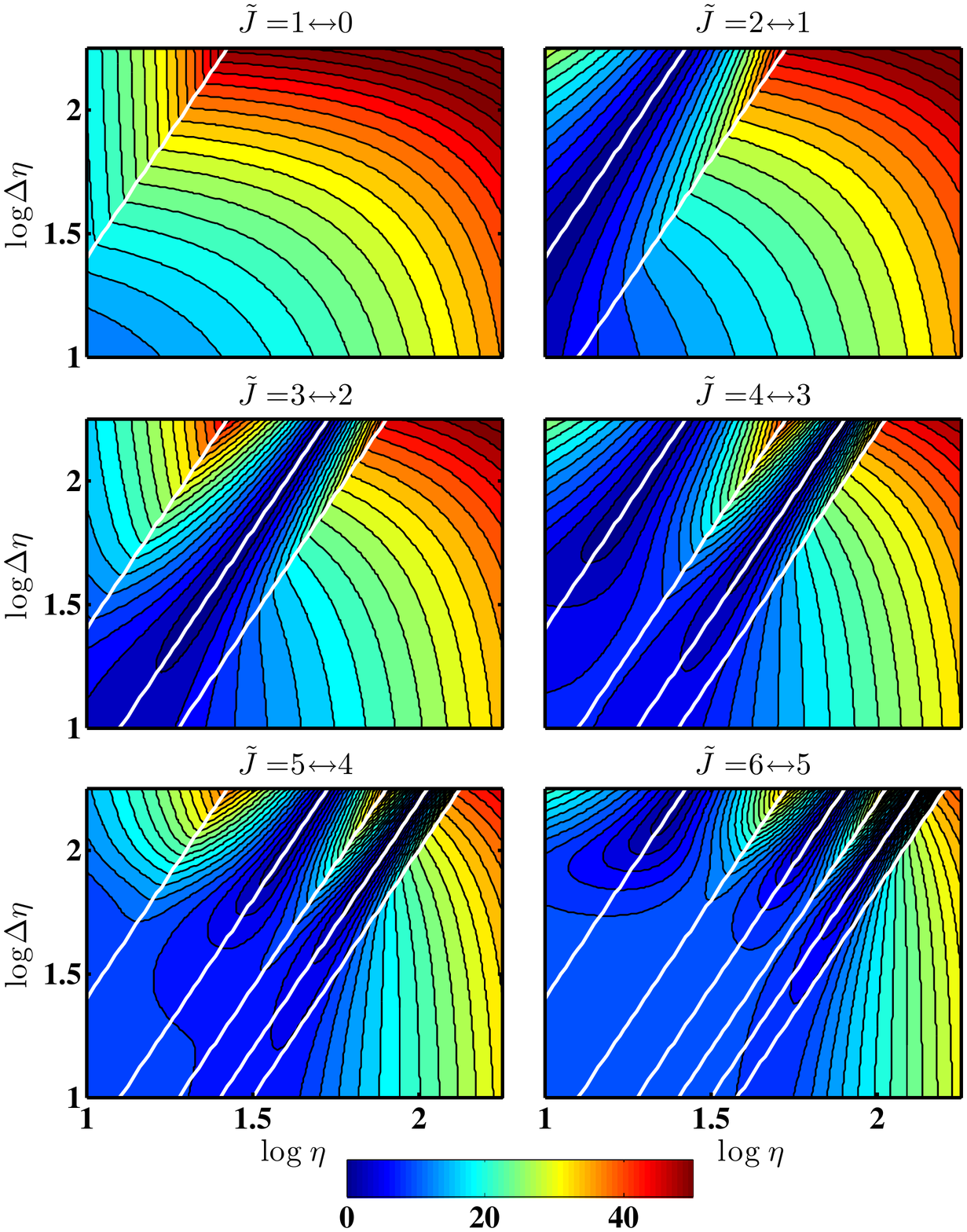}
\caption{Energy gaps, $(E_{{\tilde{J}}+1}-E_{\tilde{J}})/B$, between adjacent eigenenergy surfaces of Hamiltonian (\ref{redham})  for a linear molecule subject to an electric field. The (reduced) energy gaps are shown as functions of the parameters $\eta$ and $\Delta\eta$ that characterize the strengths of, respectively, the permanent and induced dipole interactions. White lines indicate the loci of the $k$-th order  intersection of adjacent surfaces, see Eq. (\ref{eqn:Deta-eta-simple}).}
\label{fig:gap2}
\end{figure}

In order to further visualize the topology of the energy surfaces and their intersections, we consider the energy gaps, displayed in Fig. \ref{fig:gap2}, between adjacent intersecting surfaces for the seven lowest states in the plane spanned by the interaction parameters $\eta$ and $\Delta\eta$. 
The valleys (as well as the ridges) occur along straight lines with slope two in the double-logarithmic representation of the figure, thus indicating a quadratic dependence of $\Delta\eta$ on $\eta$. 
The former ones coincide very accurately with the white lines drawn at
\begin{equation}
\Delta \eta = \frac{1}{4k^2} \eta^2
\label{eqn:Deta-eta-simple}
\end{equation}
which is equivalent to eq.~(\ref{eqn:eta-Deta-simple}) for the loci of the intersections, thereby confirming our derivation given above. Again, we see that the number of intersections an energy surface partakes in equals the adiabatic label $\tilde{J}$ of that eigenstate: While the ground state, $\tilde{J}=0$, does not exhibit any degeneracies, the first excited state, $\tilde{J}=1$, displays a first order ($k=1$) intersection with $\tilde{J}=2$ at $\Delta \eta = \eta^2/4$. In addition, the $\tilde{J}=2$ state displays a second order ($k=2$) intersection with the $\tilde{J}=3$ state at $\Delta \eta = \eta^2/16$. The $\tilde{J}=3$ state, has two more intersections with the $\tilde{J}=4$ state, one of first order and one of third order at $\Delta\eta=\eta^2/36$.

\begin{figure}[htbp]
\includegraphics[width=10cm]{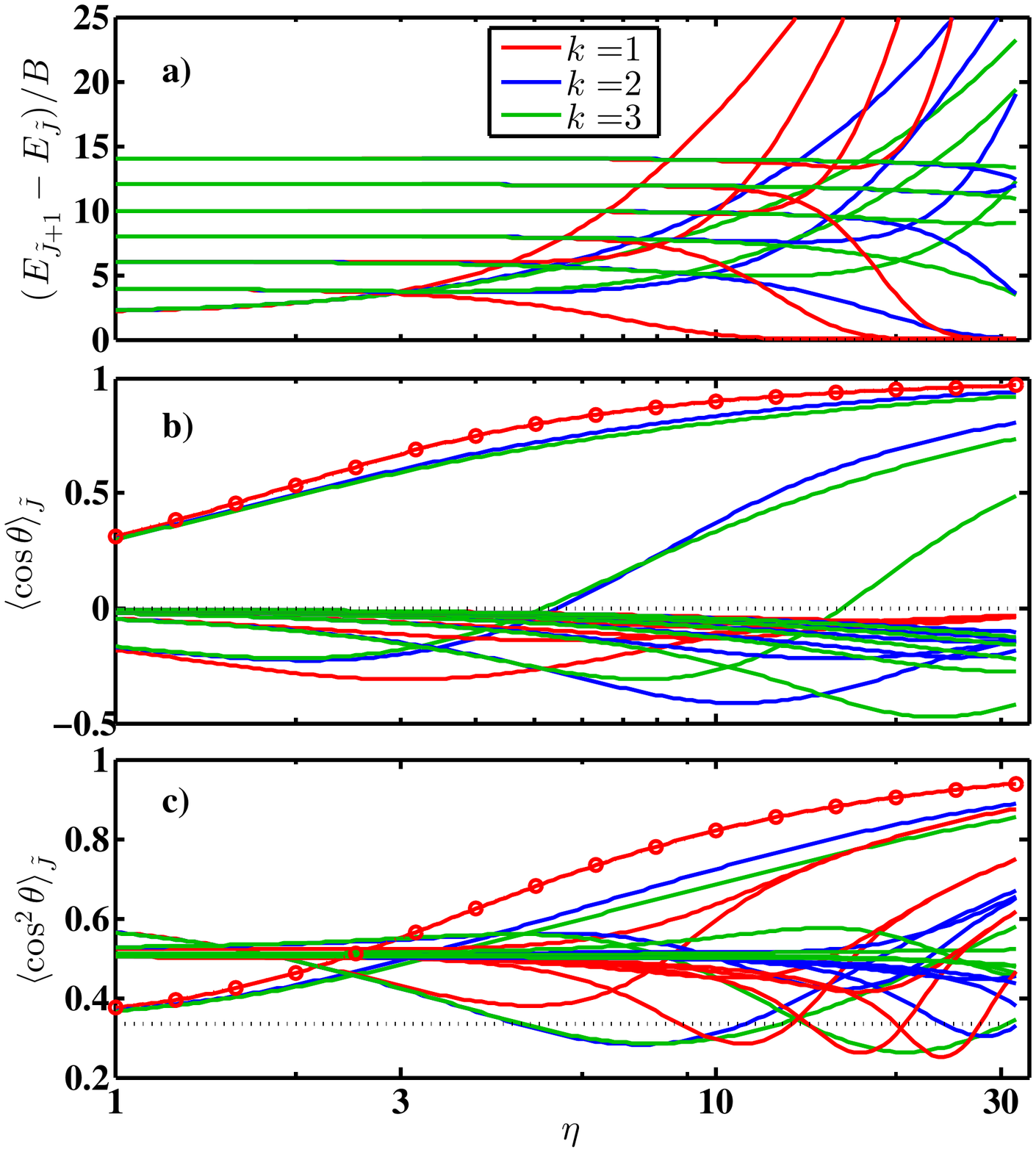}
\caption{Properties of the lowest 8 eigenstates along the lowest order intersection manifolds $\Delta\eta=\eta^2/(4k^2),\,k=1,2,3$. Top: energy gaps, $(E_{{\tilde{J}}+1}-E_{\tilde{J}})/B$, between adjacent eigenenergy surfaces of Hamiltonian (\ref{redham}). Middle: Degree of orientation, $\langle\cos\theta\rangle_{\tilde{J}}$. Bottom: Degree of alignment, $\langle\cos^2\theta\rangle_{\tilde{J}}$. The red circles superimposed on the red ground state curves in the middle and bottom panels show the analytic results obtained via supersymmetry, see Eq.~(\ref{eqn:SuSy}).}
\label{fig:properties}
\end{figure}

The $\eta$ dependence of the energy gaps between adjacent states $\tilde{J}$ and $\tilde{J}+1$  along the lowest-order intersection loci is shown in  Fig.~\ref{fig:properties}a. Beginning with $2(\tilde{J}+1)$ in the free-rotor limit, $\eta,\Delta\eta\rightarrow 0$, we find that for $k=1$ the 1--2, 3--4, ... energy gaps are decaying to nearly zero in an almost stepwise manner while the remaining ones, 0--1, 2--3, ... suddenly increase. For the $k=2$ intersection manifold, where the lowest two states remain single for all field strengths, we see a pairing of 2--3, 4--5, ...
As can also be gleaned from Fig. \ref{fig:gap2}, these drops or rises occur at lower/higher field strengths for lower/higher values of $k$ and/or $\tilde{J}$ and mirror the number of nearly-degenerate tunneling doublets that would have been generated by the induced dipole interaction alone. Conversely, at the intersection locus, the lower states may already have formed nearly-degenerate doublets while higher states are still avoiding an intersection, cf.  Fig. \ref{fig:pes}. 

\begin{figure}[htbp]
\includegraphics[width=12cm]{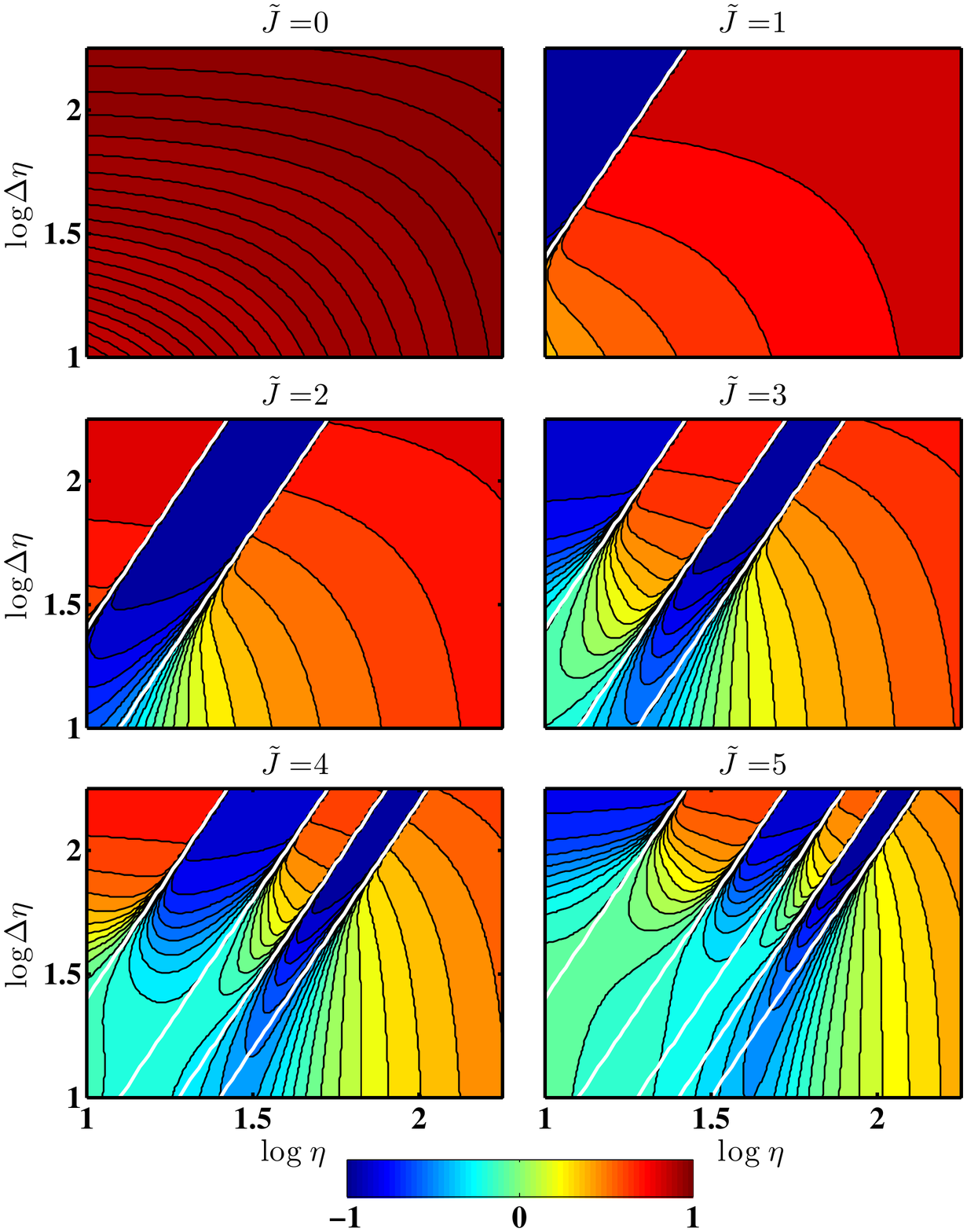}
\caption{Degree of orientation, $\langle \cos \theta \rangle_{\tilde{J}}$, for the lowest eigenstates of the Hamiltonian (\ref{redham}) for a linear molecule subject to a static electric field. White lines indicate the loci of the $k$-th order  intersections of neighboring surfaces, see Eq. (\ref{eqn:Deta-eta-simple}).}
\label{fig:ori2}
\end{figure}

\begin{figure}[htbp]
\includegraphics[width=12cm]{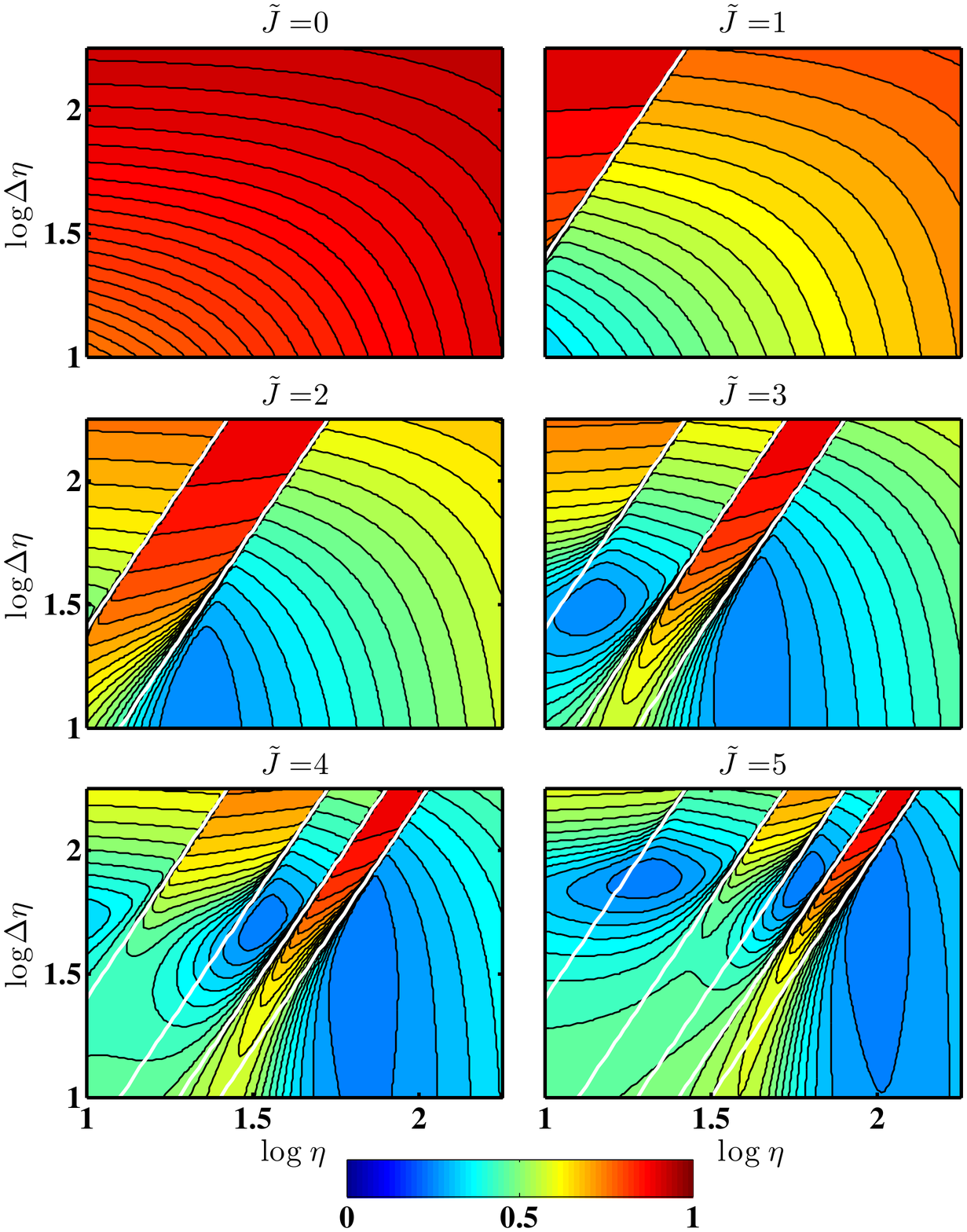}
\caption{Degree of alignment, $\langle \cos^2 \theta \rangle_{\tilde{J}}$, for the lowest eigenstates of the Hamiltonian (\ref{redham}) for a linear molecule subject to a static electric field. White lines indicate the loci of the $k$-th order  intersections of neighboring surfaces, see Eq. (\ref{eqn:Deta-eta-simple}).}
\label{fig:ali2}
\end{figure}

The directional properties of the eigenstates, as characterized by the expectation values $\langle \cos \theta \rangle_{\tilde J}$ (degree of orientation) and $\langle \cos^2 \theta \rangle_{\tilde J}$ (degree of alignment), exhibit a topology similar to that of the eigenenergies. The dependencies of the orientation and alignment cosines on the dimensionless interaction parameters $\eta$ and $\Delta\eta$ are shown, respectively, in Figs.~\ref{fig:ori2} and \ref{fig:ali2}. In addition, one--dimensional representations along the first three intersections ($k=1,2,3$) are displayed in Fig.~\ref{fig:properties}b,c.
The ground state, ${\tilde J}=0$, exhibits, at quite weak fields, high orientation and alignment, which are seen to further increase with both $\eta$ and $\Delta\eta$. 
The directionality of higher states is strongly influenced by their intersections. 
For instance, consider the first excited state, ${\tilde J}=1$. 
For $\Delta\eta>\eta^2/4$, we see a strong anti-orientation, $\langle \cos \theta \rangle_1\rightarrow-1$, together with high alignment, $\langle \cos^2 \theta \rangle_1\rightarrow 1$, see upper right panels of Figs. \ref{fig:ori2} and \ref{fig:ali2}, respectively. 
This is in keeping with the fact that this state correlates with the upper component of the lowest tunneling doublet in the limit of $\eta\rightarrow 0$, see also Fig. \ref{fig:pes}.
However, this behavior is thoroughly altered at the first-order ($k=1$) intersection where,
for $\Delta\eta<\eta^2/4$, the orientation suddenly changes its sense while the alignment is substantially reduced. 
This is connected with the fact that for $\eta\rightarrow 0$ the ${\tilde J}=1$ state intersects the lower member, ${\tilde J}=2$, of the first excited tunneling doublet at the said first order intersection.
This pattern then repeats itself  for the higher excited states. There is always an upper doublet member (anti-oriented for sufficiently large $\Delta\eta$ and for $\eta\rightarrow 0$) crossing a lower doublet member (oriented for sufficiently large $\Delta\eta$ and for $\eta\rightarrow 0$) at each of the intersections, see Fig. \ref{fig:ori2}.
As a result, the ${\tilde J}$--th state in the combined fields exhibits ${\tilde J}$ sign changes of the orientation cosine, and these sign changes occur abruptly at the first ${\tilde J}$ intersections as given by Eq. (\ref{eqn:Deta-eta-simple}).
For the same reason, at these loci, the degree of alignment is found to be almost discontinuous as well, see Fig. \ref{fig:ali2}. 
We note that for the higher doublets, the directionality tends to vanish as those states are unbound and their orientation and alignment approaches that of an (isotropic) free rotor.

Fig.~\ref{fig:properties}b attests, in addition, that at the loci of the first-order intersections, $k=1$, the ground-state orientation is always positive, approaching unity at strong fields while all other states shown exhibit a weak anti-orientation which tends to vanish at strong fields. For higher-order intersections, $k>1$, the first $k$ states also exhibit a nearly ``perfect orientation'' with increasing field strength while the remaining states are always anti-oriented with a vanishing orientation at high fields. The weak anti-orientation along the intersections seams seen for $\tilde{J}\ge k$ in Fig. \ref{fig:properties}b implies that the sign changes of the orientation cosine do not occur exactly at the intersection loci (white lines in Fig. \ref{fig:ori2}) but are slightly shifted towards the ``foothills,'' cf. Fig. \ref{fig:ori2}.

Fig.~\ref{fig:properties}c details the alignment cosines along the intersection seams for the eigenstates considered. Except for the ground state, the alignment is first decreasing, exhibiting anti-alignment ($\langle \cos^2 \theta \rangle<\frac{1}{3}$) before rising again and approaching unity at strong fields. Neither here do the almost discontinuous changes of the alignment cosine coincide with the intersection loci (white lines in Fig. \ref{fig:ali2}) exactly but are slightly shifted away from the ridges. 

\section{Applications and prospects}

\begin{figure}[htbp]
\centering
\includegraphics[width=20cm]{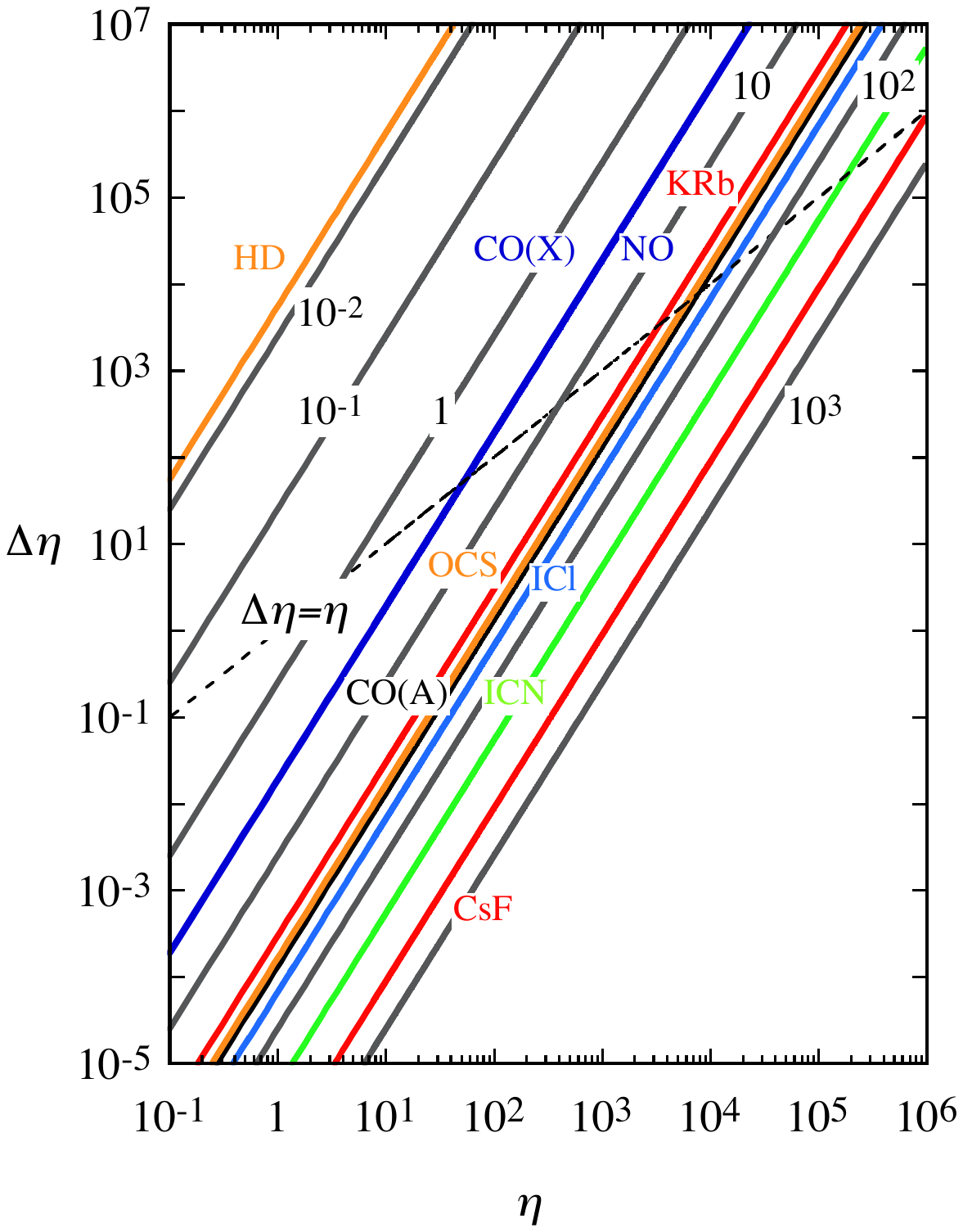}
\caption{\label{fig:Plot-eta-Deta} {\small 
Nomogram for the dependence of the induced dipole interaction parameter $\Delta \eta$ on the permanent dipole interaction parameter $\eta$ as given by eq. (\ref{eta-Deta}) for the set of molecules listed in Table \ref{table:parameters}. Also shown are the $\Delta \eta=\eta$ line (dashed), above which the induced-dipole interaction exceeds the permanent dipole interaction, and the dependence of $\Delta \eta$ on $\eta$ for a sampling of integer values of $k$ (which label the grey lines).}}
\end{figure}

If both the permanent and induced dipole interactions arise from the same field $\boldsymbol{\varepsilon}_{1}=\boldsymbol{\varepsilon}_{2}=\boldsymbol{\varepsilon}$, it follows from eq. (\ref{defparam}) that the ratio of the combined permanent and induced electric dipole interaction parameters  is fixed for a given molecule with a body-fixed permanent electric dipole moment, $\mu$, polarizability anisotropy, $\Delta \alpha$, and rotational constant, $B$,
\begin{equation}
\frac{\Delta \eta}{\eta^2}=\frac{\Delta \alpha B}{2 \mu^2}  \label{eta-Deta}.
\end{equation}
Figure \ref{fig:Plot-eta-Deta} displays this dependence of the induced dipole interaction parameter $\Delta \eta$ on the permanent dipole interaction parameter $\eta$ for the molecules listed in Table \ref{table:parameters}. Note that the higher the value of the $\frac{\Delta \alpha B}{2 \mu^2}$ parameter, the more easy it is to reach the regime where the induced-dipole interaction exceeds the permanent dipole interaction. This regime arises above the $\Delta \eta=\eta$ line, also shown in Fig. \ref{fig:Plot-eta-Deta}.

This observation puts our main result --  Eq. (\ref{eqn:Deta-eta-simple}) for the loci of the Stark energy intersections -- into a new perspective: 
First of all, the quadratic dependence of $\Delta\eta$ on $\eta$  is exactly what obtains for a fixed ratio $\frac{\Delta \alpha B}{2 \mu^2}$ pertaining to a given molecule subject to an electric field $\boldsymbol{\varepsilon}_{1}=\boldsymbol{\varepsilon}_{2}=\boldsymbol{\varepsilon}$, cf. Fig. \ref{fig:Plot-eta-Deta}. It follows that
the quantum dynamics induced by an electric field $\boldsymbol{\varepsilon}$ (of arbitrary time dependence) will be also constrained to lines with slope two in a double-logarithmic representation of the ($\eta,\Delta\eta$) plane (such as the one in Fig. \ref{fig:gap2}), i.e., parallel to the intersection loci.
This motivates assigning the index $k$ to molecules according to
\begin{equation}
k = \frac{\eta}{2\sqrt{\Delta\eta}}
\label{eqn:k}
\end{equation}
which is listed in Table~\ref{table:parameters} for the choice molecules. The dependence of  $\Delta \eta$ on $\eta$ for selected values of $k$ is included in Fig. \ref{fig:Plot-eta-Deta}.
Depending on whether the $k$ index is (nearly) integer or, say, (nearly) half-integer, the resulting dynamics will be qualitatively different. 
Whereas in the former case, an increase in the field strength will drive the system into the intersections, in the latter case the intersections will be avoided. We note that the ground state never partakes in any intersections.

Yet another intriguing feature of the intersections is their apparent connection to supersymmetry. At the loci of the first order intersections, $\Delta\eta=\eta^2/4$, see eq. (\ref{eqn:Deta-eta-simple}) with $k=1$, the eigenproperties of Hamiltonian  (\ref{redham}) were previously derived analytically  in closed form via the apparatus of supersymmetric quantum mechanics (SUSY QM) \cite{LemMusKaisFriPRA2011,LemMusKaisFriNJP2011}. The analytic solution was obtained for  a class of states (stretched states, with $M=\tilde{J}$) for a particular ratio of the interaction parameters, namely for 
\begin{equation}\label{susy}
\frac{\Delta\eta}{\eta^2}=\frac{1}{4(M+1)^2}
\end{equation}
which yields $\Delta\eta=\eta^2/4$ for $M=0$. We emphasize that both $\eta$ and $\Delta \eta$ must be nonzero in order for the analytic solutions to exist, which means that these are not available for either the permanent or induced dipole interaction acting alone.

As an example, we list the analytic SUSY results for the energy, orientation, and alignment of the ground state along the loci of the first order intersection, $\Delta\eta=\eta^2/4$, 
\begin{eqnarray}
\frac{E_0}{B} &=& -\frac{\eta^2}{4}=-\Delta\eta \nonumber \\
\langle \cos \theta \rangle_0 &=& \coth \eta - \frac{1}{\eta} \nonumber \\
\langle \cos^2 \theta \rangle_0 &=& 1 + \frac{2}{\eta^2} - \frac{2 \coth \eta}{\eta}
\label{eqn:SuSy}
\end{eqnarray}
all of which are reproduced quantitatively by our numerical results; the latter two are shown in Fig. \ref{fig:properties}b,c. 

Hence our present work provides an additional insight, namely that the condition for the existence of an analytic solution for the ground state coincides with the condition for the intersection loci of the first and second excited states of the underlying combined-fields Hamiltonian.

\newpage

\bibliographystyle{ieeetr} 
\def\refname{}

\vspace{-1cm}
\bibliography{CombFixed}

\end{document}